\documentclass[aps,prl,twocolumn,groupedaddress,showpacs,tightenlines]{revtex4}
\begin{document}

\title{Generalization of Boltzmann Equilibration Dynamics}

\author{Travis Sherman and Johann Rafelski}
 \affiliation{Physics Department, University of Arizona, Tucson, Arizona 85721}

\date{\small July 25, 2002}

\begin{abstract}
We propose a novel approach in the study of transport phenomena in dense systems or systems with long range interactions where multiple particle interactions must be taken into consideration.  Within Boltzmann's kinetic formalism, we study the influence of other interacting particles in terms of a random distortion of energy and momentum conservation occurring when multi-particle interactions are considered as binary collisions. Energy and momentum conservation still holds exactly but not in each model binary collision.  We show how this new system differs from the Boltzmann system and we note that our approach naturally explains the emergence of Tsallis-like equilibrium statistics in physically relevant systems in terms of the long since neglected physics of interacting and dense systems.
\end{abstract}
\maketitle
The Boltzmann equation \cite{Bol64} for a spatially independent system of particles describes the evolution of the one particle momentum $\vec{p}$ distribution $f(\vec{p},t)$ of a ``foreground'' particle with mass $m$ subject to interactions (here assumed to be number conserving) with $N$ ``background'' particles $j=1,\ldots,N$ with masses $m_1,m_2,\ldots,m_{N}$.  In classical, non-relativistic regimes, the Boltzmann equation takes the form
\begin{equation}
\frac{\partial f(\vec{p},t)}{\partial t}\!=\!\!\int\!\!\left[ W(\vec{p}\,',\vec{p})f(\vec{p}\,',t)-\!W(\vec{p},\vec{p}\,')f(\vec{p},t)\right]d^3\!\vec{p}\,'\!\!, \label{classicalboltzmanneq}
\end{equation}
where we have suppressed a possible time dependence in the transition rate $W(\vec{p},\vec{p}\,')$, the rate per unit time of the foreground particle making a momentum transition from $\vec{p}$ to $\vec{p}\,'$ due to interactions with the background particles.

In applications, the Boltzmann equation is restricted to sufficiently rarefied systems with short range interactions so that only two particle interactions (two body or binary collisions) are incorporated into the transition rate \cite{Uhl73}.  For these systems, the transition rate is given by \cite{Kam81}
\begin{eqnarray}
&&W(\vec{p},\vec{p}\,')\!\!=\!\!\!\sum_j\Big[\!\int\!\!d^3\!\vec{q_j}\!\!\int\!\!d^3\!\vec{q}_j\!\,'\, \delta^3(\vec{p}+\vec{q_j}-\vec{p}\,'-\vec{q}_j\!\,')\nonumber\\
&&\times\delta(\text{E}_p+\text{E}_{q_j}-\text{E}_{p\,'}-\text{E}_{q_j'})(\sigma v)_j b_j(\vec{q}_j,t)\Big],\label{TransitionRate1}
\end{eqnarray}     
where the energies are given by $\text{E}_p=|\vec{p}\,|^2\!/2m$ and $\text{E}_{q_j}=|\vec{q_j}|^2\!/2m_j$, $b_j(\vec{q_j},t)$ is the momentum $\vec{q_j}$ distribution for background particle $j$, and $(\sigma v)_{j}$ is the differential cross section for the scattering of the foreground particle with a background particle $j$.  

No satisfactory generalization of the transition rates has been given to describe transport phenomena in dense systems or systems with long range interactions, that is, in systems where multiple particle interactions must be taken into consideration \cite{Coh73}.  The natural generalization of the transition rates is to assume that the transition rate is proportional to the sum over collections of background particles of \textbf{1)} the expectation of a collision between the foreground particle and a collection of background particles and \textbf{2)} the expectation of the conservation of energy and momentum constraint.    However, there is no clear way of systematically generalizing the Boltzmann equation by adding successive multi-particle-interaction correction terms (which are generally divergent). Attempts to generalize the Boltzmann equation for even ``moderately'' dense systems have  largely been abandoned.

In dense systems or systems with long range interactions, for the transition $\vec{p}\rightarrow\vec{p}\,'$ of the foreground particle to be physically possible in an interaction, we need the initial and final momenta and energies to satisfy
\begin{eqnarray} 
\vec{p}+\vec{q}_1+\ldots+\vec{q}_{N}&=&\vec{p}\,'+\vec{q}\,'_1+\ldots+\vec{q}\,'_{N}\label{constraint1}\\
\text{E}_p+\text{E}_{q_1}+\ldots+\text{E}_{q_{N}}&=&\text{E}_{p'}+\text{E}_{q_1'}+\ldots+\text{E}_{q_{N}'}\label{constraint2}
\end{eqnarray}
by the conservation of momentum and energy.  Singling out the dominant role that a single background particle $j$ plays in making the transition physically possible, we write the above constraints as
\begin{eqnarray}
\vec{p}+\vec{q}_j&=&\vec{p}\,'+\vec{q}_j\,'+\vec{\epsilon}\\
\text{E}_p+\text{E}_{q_j}&=&\text{E}_{p'}(2\gamma-1)+\text{E}_{q_j'}.
\end{eqnarray}
The latter form of the constraints appears arbitrary. However, when combined with Eqs.~(\ref{constraint1}) and~(\ref{constraint2}), it serves as a defining relation for the variables $\vec{\epsilon}$ and $\gamma$.  The above form is chosen here as such for mathematical convenience.  

Thus, in dense systems or systems with long range interactions, for the transition to be ``due'' to an interaction with particle $j$, we need a foreground-background binary interaction as before, but with modified energy and momentum constraints.  The transition rates for dense systems or systems with long range interactions are then given by simply replacing the $\delta$-functions appearing in Eq.~(\ref{TransitionRate1}) according to:
\begin{widetext}
\begin{eqnarray}
\delta^3(\vec{p}+\vec{q_j}-\vec{p}\,'-\vec{q_j}\,')&\rightarrow &\langle \langle \delta^3(\vec{p}+\vec{q_j}-\vec{p}\,'-\vec{q_j}\,'-\vec{\epsilon})\rangle \rangle_{\vec{\epsilon}}\\ 
\delta(\text{E}_p+\text{E}_{q_j}-\text{E}_{p'}-\text{E}_{q_j'})&\rightarrow&\langle \langle \delta[\text{E}_p+\text{E}_{q_j}-\text{E}_{p'}(2\gamma-1)-\text{E}_{q_j'}] \rangle \rangle_{\gamma}.
\end{eqnarray}
\end{widetext}
So far, we have pushed all of the complexity of computing multiple scattering transition rates into computing the distributions of the random vector $\vec{\epsilon}$ and the random variable $\gamma$.  However, the form of the transition rate is now familiar and many aspects of the distributions are immediately clear from their definitions.  As the density of a system and/or the range of interactions decrease, $\langle\langle \gamma \rangle\rangle_{\gamma} \rightarrow 1$, $\langle\langle \vec{\epsilon}\rangle\rangle_{\vec{\epsilon}} \rightarrow\vec{0}$ and $\text{Var}(\gamma)\rightarrow 0$, $\text{Var}(\vec{\epsilon})\rightarrow\vec{0}$ for only two particle collisions are increasingly present.  Clearly, our transition rates simplify to the usual binary transition rates for rarefied systems with short range interactions and thus, incorporate the successes of the Boltzmann equation in describing transport phenomena in these systems.  

However, as the density of a system and/or as the range of interactions increase, $\text{Var}(\gamma)$ and $\text{Var}(\vec{\epsilon})$ increase.  For such physically relevant systems, the success of our proposed generalization to the Boltzmann equation will be made evident by indicating how our generalization determines generalized equilibration dynamics which differ from canonical equilibration dynamics and how our generalization very naturally reproduces experimentally measured equilibrium transport phenomena.             

In an interaction involving a foreground particle and a background particle $j$ we can interpret the presence of other interacting background particles in dense systems or systems with long range interactions as providing a mechanism for carrying off or providing the additional momentum and/or energy necessary to make transitions physically possible.  We can also interpret the above transition rate as the usual transition rate in a system in which only binary collisions occur which do not necessarily conserve kinetic energy and momentum, as is the case for inelastic collisions.  

Not only do these two interpretations aid us in understanding the nature of the $\gamma$ and $\vec{\epsilon}$ distributions, but we see that by extending our original definitions of $\vec{\epsilon}$ and $\gamma$, the form of our transition rates is applicable to a very large class of systems: with elastic and/or inelastic interactions (here defined generally to incorporate all interactions which do not conserve energy and momentum), with low or high number densities, with short and/or long range interactions, and with any mechanisms which can carry off or provide additional momentum and energy to particles.  

Indeed, for general systems, the calculation of the $\vec{\epsilon}$ and $\gamma$ distributions has become even more complicated.  Before proceeding, it is helpful to note that the expectation of the parameters of our generalization appearing in the delta functions of Boltzmann's formalism has the effect of widening or broadening the delta functions.  Thus, the existence of our parameters with some fluctuation is guaranteed in that no exact delta function exists in nature. As such, the importance of our proposed generalization to physically relevant systems depends on the extent to which systems exhibit fluctuations in the parameters $\gamma$ and $\vec{\epsilon}$.  

It is natural to expect that systems with large number densities, long range interactions, inelastic interactions, and other such mechanisms (such as quantum mechanical interactions) exhibit sufficient fluctuations to, and indeed as they are known to, require a generalized formalism.  Thus, to verify the necessity of generalizing Boltzmann's formalism to a system requires extensive theoretical calculations and/or precise experimental measurements of the $\gamma$ and $\vec{\epsilon}$ parameters of that system.  One such theoretical calculation has already been given by H. Haug and C. Ell~\cite{Hau92}, whose derivation of a semiclassical Boltzmann equation for Coulomb quantum kinetics in a dense electron gas suggests, in the limit of completed collisions, an asymptotic and approximate distribution of gamma given by a peaked distribution about $\gamma=1$:
$
 f_\gamma(\gamma)=2\left(\Gamma/2\text{E}_p'\right)
 /  
\left[(\gamma-1)^2+\left(\Gamma/2\text{E}_p'\right)^2\right],
$
 where $\Gamma$ is the sum of the collision dampening coefficients.

In the absence of other such calculations or measurements of the distributions of $\gamma$ and $\vec{\epsilon}$ in various physical systems, we may still proceed in the verification of our theory since the precise form of our proposed generalization can be verified by comparing the consequences of  our proposal with well established and currently unexplained phenomena.  For instance, we can  study transport phenomena within our formalism which result from certain classes of $\vec{\epsilon}$ and $\gamma$ distributions which we expect to be present in many physically relevant systems.  

In many systems, we expect $\vec{\epsilon}$ and $\gamma$ to be peaked distributions about $\vec{\epsilon}=\vec{0}$ and $\gamma=1$ and approximately independent of the incoming and outgoing momenta and energies (so that we interpret the influence of other interacting particles, unaccounted interaction processes and mechanisms as producing a random noise distortion of the conservation of energy and momentum constraints).  In particular, we might expect $\vec{\epsilon}$ to have a spherically symmetric distribution about $\vec{\epsilon}=\vec{0}$.  Moreover, since the family of Gamma distributions has a rich variety of shapes and can approximate many classes of distributions, in many systems we expect the peaked distribution of $\gamma$ to be well approximated by a Gamma distribution with parameters $\alpha$ and $\lambda$, chosen so that the distribution is peaked about $\gamma=1$:
\begin{equation}\label{distgamma} 
f_\gamma(\gamma)= 
{\lambda(\lambda\gamma)^{\alpha-1}e^{-\lambda\gamma}}/{\Gamma(\alpha)},
\end{equation}
 where the average and variance of $\gamma$ are $\text{Avg}(\gamma)\equiv\langle\langle \gamma \rangle\rangle_{\gamma}={\alpha}/{\lambda}$ and $\text{Var}(\gamma)\equiv\langle\langle \gamma^2 \rangle\rangle_{\gamma}-\langle\langle \gamma \rangle\rangle_{\gamma}^2={\alpha}/{\lambda^2}$.  

The form  of the distribution of $\gamma$ as a Gamma distribution is chosen here for mathematical convenience and to analytically recover the precise form of Tsallis' proposed equilibrium distribution in the following example, but we emphasize that any peaked distribution of $\gamma$ about $\gamma=1$ will yield a Tsallis-like distribution, i.e., a deviation from the Boltzmann equilibrium distribution.  Numerical calculations with the distribution of gamma suggested by the work of Haug and Ell resulted in Tsallis-like equilibrium distributions for a large range of $\Gamma$ and the Boltzmann equation was recovered as the width $\Gamma\rightarrow0$.

To analytically illustrate some immediate results of our proposed generalization, we consider a system of $N$ classical background particles denoted by $j=1,...,N$ with mass $m_j$.  Suppose further that the interactions are completely elastic and that the number density of the system and the range of the interactions are sufficiently small to guarantee that the momentum distribution of the background particle $j=1,...,N$ approaches the Boltzmann equilibrium distribution with a common temperature $T$ (measured in units of energy): $ b^{\text{eq}}_j(q_j)=C_j\,\exp\left(-q_j^2/2m_jT\right)$ where $C_j$ is a normalization constant.  After the $N$ particles are sufficiently equilibrated, we inject a foreground particle with mass $m\gg m_j$ for all background particles $j=1,...,N$ and an initial distribution $f(\vec{p},t=0)$.  Suppose that multi-particle collisions between the foreground and background particles occur (for instance, as the diameter of the foreground particle is much larger than the diameter of the background particles or if the range of interaction is larger for a foreground-background interaction than for a background-background interaction) and/or that the interactions are inelastic.  Thus, the distributions of $\gamma$ and $\vec{\epsilon}$ exhibit variation.  Further, suppose $(\sigma v)_j$ is constant as in the case of hard sphere interactions.  We also suppose that the background particles are sufficient in extent (i.e., the background is an ideal heat bath) so that we may assume that the background approximately remains in equilibrium during the evolution of the foreground particle.  Therefore, the time dependent distribution $f(\vec{p},t)$ satisfies Eq.~(\ref{classicalboltzmanneq}), where the transition rate is 
\begin{eqnarray}
W(\vec{p},\vec{p}\,')\!\!&=&\!\!\sum_j\!\!\int\!d^3\!\vec{q_j}\int\!d^3\!\vec{q}_j\,' \langle\langle\delta^3(\vec{p}+\vec{q_j}-\vec{p}\,'-\vec{q}\,'_j-\vec{\epsilon})\rangle\rangle_{\vec{\epsilon}}\nonumber\\
&&\times\,\langle\langle\delta[\text{E}_p+\text{E}_{q_j}-\text{E}_{p\,'}(2\gamma-1)-\text{E}_{q_j'}]\rangle\rangle_{\gamma}\nonumber\\
&&\times\,(\sigma v)_j\,\,b^{\text{eq}}_j(\vec{q}_j). \label{firstrate}
\end{eqnarray}
Interchanging the order of integration and using the $\delta^3$-function to integrate out $\vec{q}_j\,'$ (so that $\vec{q}_j\,'=\vec{p}+\vec{q}-\vec{p}\,'-\vec{\epsilon}$), we obtain
\begin{eqnarray}
&&W(\vec{p},\vec{p}\,')\!\!=\!\Bigg\langle\!\!\Bigg\langle\!\!\sum_j\!\!\int\!q_j^2\,d q_j\,d(\cos\theta_j)\,d\phi_j\,\delta\Big(\frac{p^2+p'^2}{2m}\nonumber\\
&&-\gamma\frac{p'^2}{m}-\frac{|\vec{p}-\vec{p}\,'-\vec{\epsilon}\,|^2}{2m_j}-\frac{|\vec{p}-\vec{p}\,'-\vec{\epsilon}\,|q_j\cos\theta_j}{m_j}\Big)\nonumber\\
&&\times \,(\sigma v)_j\,\,C_j\,e^{\displaystyle-\frac{q_j^2}{2m_jT}}\Bigg\rangle\!\!\Bigg\rangle_{\vec{\epsilon},\gamma}
\end{eqnarray}
where we have introduced spherical coordinates $(q_j,\theta_j,\phi_j)$ for the integration over $\vec{q}_j$.  Note that we have chosen $\theta_j$ to measure the angle between $\vec{q}$ and $\vec{p}-\vec{p}\,'-\vec{\epsilon}$ so that the  integration over $\phi_j$ is trivial and yields a factor of $2\pi$.  We can now integrate over $\cos\theta_j$ using the remaining delta function.  A required transformation introduces a factor $(|\vec{p}-\vec{p}\,'-\vec{\epsilon}\,|\,\,q_j)^{-1}m_j$ and the single zero of the delta function uniquely determines $\cos\theta_j$:
\begin{equation}
\cos\theta_j\!\!=\!\!\left(\frac{p^2+p'^2}{2m}-\gamma\frac{p'^2}{m}-\frac{|\vec{p}-\vec{p}\,'-\vec{\epsilon}|^2}{m_j}\right)\!\!\frac{m_j}{|\vec{p}-\vec{p}\,'-\vec{\epsilon}\,|q_j}.
\end{equation}
The result is
\begin{eqnarray}
W(\vec{p},\vec{p}\,')\!\!=\!\!\Bigg\langle\!\!\Bigg\langle\!\!\sum_j\!\!\int\!\!dq_j\!\frac{2\pi q_j m_j (\sigma v)_j C_j }{|\vec{p}-\vec{p}\,'-\vec{\epsilon}|} e^{\displaystyle-\frac{q_j^2}{2m_jT}}\!\Bigg\rangle\!\!\Bigg\rangle_{\vec{\epsilon},\gamma}
\end{eqnarray}
where the integration over $q_j$ is taken over all positive $q_j$ which satisfy $-1\leq\cos\theta_j\leq 1$.  Solving this constraint implies that 
\begin{eqnarray}
q_j^2\geq\frac{|\vec{p}-\vec{p}\,'-\vec{\epsilon}|^2}{4}+\left(p'^2\gamma-\frac{p^2+p'^2}{2}\right)\!\frac{m_j}{m}
\end{eqnarray}
to first order in $m_j/m$.  The remaining integral over $q_j$ is trivial and one obtains
\begin{eqnarray}
&&W(\vec{p},\vec{p}\,')\!=\!\sum_j\,2\pi m_j^2T\,(\sigma v)_j\,C_j\,e^{\displaystyle \frac{\text{E}_p+\text{E}_{p'}}{2T}}\nonumber\\
&&\times \left\langle\!\!\left\langle \frac{e^{\displaystyle-\frac{|\vec{p}-\vec{p}\,'-\vec{\epsilon}|^2}{8m_jT}}}{|\vec{p}-\vec{p}\,'-\vec{\epsilon}|}\right\rangle\!\!\right\rangle_{\vec{\epsilon}}
\left\langle\!\!\left\langle e^{\displaystyle-\frac{\gamma\,\text{E}_{p'}}{T}} \right\rangle\!\!\right\rangle_{\gamma}.\label{reducedEq}
\end{eqnarray}    
The final form of transition rate indicates that the distribution of $\vec{\epsilon}$ determines the rate of equilibration, while the distribution of $\gamma$ determines the shape of the resulting equilibrium.  Now, if $\vec{\epsilon}$ is symmetrically distributed about $\vec{\epsilon}=\vec{0}$ (as we suggested above), we have that 
\begin{equation}
\left\langle\!\!\left\langle \frac{e^{\displaystyle-\frac{|\vec{p}-\vec{p}\,'-\vec{\epsilon}|^2}{8 m_jT}}}{|\vec{p}-\vec{p}\,'-\vec{\epsilon}|}\right\rangle\!\!\right\rangle_{\vec{\epsilon}}
=\left\langle\!\!\left\langle \frac{e^{\displaystyle-\frac{|\vec{p}-\vec{p}\,'+\vec{\epsilon}|^2}{8 m_jT}}}{|\vec{p}-\vec{p}\,'+\vec{\epsilon}|}\right\rangle\!\!\right\rangle_{\vec{\epsilon}}.\label{identity}
\end{equation}
Multiplying Eq.~(\ref{reducedEq}) by $\langle\langle \exp(-\gamma E_p/T) \rangle\rangle_\gamma$, interchanging $\vec{p}\leftrightarrow\vec{p}\,'$ and making use of Eq.~(\ref{identity}), we see that
\begin{equation}
W(\vec{p}\,',\vec{p})\left\langle\!\!\left\langle e^{\displaystyle-\frac{\gamma\text{E}_{p'}}{T}} \right\rangle\!\!\right\rangle_{\gamma}=W(\vec{p},\vec{p}\,') \left\langle\!\!\left\langle e^{\displaystyle-\frac{\gamma\text{E}_{p}}{T}} \right\rangle\!\!\right\rangle_{\gamma}\label{identity2}.
\end{equation}
Note that this is a detailed balance equation \cite{Kam81}, implying
\begin{equation}
f^{\text{eq}}(\vec{p})=C\left\langle\!\!\left\langle e^{\displaystyle-\frac{\gamma\text{E}_{p}}{T}} \right\rangle\!\!\right\rangle_{\gamma}\label{dist}
\end{equation}
is a time-independent solution of Eq.~(\ref{classicalboltzmanneq}), where $C$ is a normalization constant.  This is easily verified by substituting $f^{\text{eq}}(\vec{p})$ in Eq.~(\ref{dist}) for $f(\vec{p},t)$ in Eq.~(\ref{classicalboltzmanneq}) and by making use of Eq.~(\ref{identity2}).  If the distribution of $\gamma$ is well approximated by a Gamma distribution with parameters $\alpha$ and $\lambda$ (as suggested above), we find that
\begin{equation}
f^{\text{eq}}(\vec{p})=C\!\left(1+\frac{\text{E}_p}{T}\lambda\right)^{-\alpha}.
\end{equation}
Thus, interpreting the Tsallis \cite{Tsa88} nonextensitivity parameter $q_{\text{T}}$ as
\begin{equation}
q_{\text{T}}=1+\frac{1}{\alpha}=\frac{\langle\langle \gamma^2 \rangle\rangle_{\gamma}}{\langle\langle \gamma \rangle\rangle_{\gamma}^2}
\end{equation}
and the inverse temperature of the foreground particle $\beta_{\text{F}}=1/T_{\text{F}}$ as $ \beta_{\text{F}}=\frac{\alpha}{\lambda}\beta=\langle\langle \gamma \rangle\rangle_{\gamma} \beta$,
we have that the equilibrium distribution of the foreground
particle is a Tsallis equilibrium distribution with
nonextensitivity parameter $q_{\text{T}}$, inverse foreground
temperature $\beta_F=1/T_{\text{F}}$, and normalization constant $C$:
\begin{equation}
f^{\text{eq}}(\vec{p})=C \left[1-\beta_{\text{F}}(1-q_{\text{T}})\text{E}_p\right]^{\frac{1}{1-q_{\text{T}}}}.
\end{equation}
The Tsallis equilibrium distribution is often successfully used to model equilibrium distributions exhibiting power-tail behavior.  Furthermore, Tsallis-like distributions are extensively measured in many physical systems, especially in systems with long range interactions and in turbulent flows~\cite{Bec01}.  

We note that our approach differs from \cite{Lim01} in that our ``natural'' generalization does not up-front introduce  exact Tsallis equilibration and, moreover, we retain the intuitive statistical form of Boltzmann's molecular chaos hypothesis, transport equation, and transition rates.  Our proposal compliments the approach taken by C. Beck~\cite{Bec2}, which generates Tsallis equilibrium statistics by considering fluctuations in the parameters of the Langevin equation.  We  incorporate the approach taken by G. Wilk and Z. Wlodarczyk~\cite{Wil00}, which generates Tsallis equilibrium statistics from fluctuations in background temperatures in the Boltzmann exponential factor, by incorporating these fluctuations into our distribution of $\gamma$. 

Concluding, we have proposed a  generalization of Boltzmann equilibration dynamics to model transport phenomena in general, non-equilibrium systems.  Our generalization resulted from our interpretation of the influence of other interacting particles as a random distortion of energy and momentum conservation occurring when multiple interactions are considered as binary collisions. As it turns out  this also explains the appearance of Tsallis distribution in Fokker-Planck dynamics \cite{Wal00}, where local energy conservation was not maintained. 

In outlook, we noted that our formalism is applicable to many different and large classes of systems: with elastic and/or inelastic interactions, with low or high number densities, with short and/or long range interactions, and with any mechanisms which can carry off or provide additional momentum and energy to particles.  We also indicated how the result of our proposed generalization can be determined in systems by studying classes of reasonable distributions for the proposed parameters of our generalization.  Although we have only considered a simple example where we could easily and explicitly determine some immediate results, our approach is easily extended to more general systems.  We already have similar results for a relativistic, quantum mechanical foreground particle scattering in a thermal background of light relativistic particles, and it is clear that the general situation is similar.  
  
\vskip 0.3cm
\noindent{\bf Acknowledgments:}
Work supported in part by a grant from the U.S. Department of
Energy,  DE-FG03-95ER40937.  Travis Sherman was in part 
supported by a UA/NASA Space Grant Undergraduate Research
Internship Program.


\end{document}